# Linking Microstructural Evolution and Macro-Scale Friction Behavior in Metals

N. Argibay[1], M. Chandross[1,*], S. Cheng[2], J.R. Michael[1]

[1] *Materials Science and Engineering Center, Sandia National Laboratories, Albuquerque, NM 87185*
[2] *Department of Physics, Center for Soft Matter and Biological Physics and Macromolecules Innovation Institute, Virginia Polytechnic Institute and State University, Blacksburg, VA 24061*

*N. Argibay and M. Chandross contributed equally to this work.*

*\* Corresponding author e-mail address: mechand@sandia.gov*

## Abstract

A correlation is established between the macro-scale friction regimes of metals and a transition between two dominant atomistic mechanisms of deformation. Metals tend to exhibit bi-stable friction behavior -- low and converging or high and diverging. These general trends in behavior are shown to be largely explained using a simplified model based on grain size evolution, as a function of contact stress and temperature, and are demonstrated for pure copper and gold. Specifically, the low friction regime is linked to the formation of ultra-nanocrystalline surface films (10 to 20 nm), driving toward shear accommodation by grain boundary sliding. Above a critical combination of stress and temperature -- demonstrated to be a material property -- shear accommodation transitions to dislocation dominated plasticity and high friction. We utilize a combination of experimental and computational methods to develop and validate the proposed structure-property relationship. This quantitative framework provides a shift from phenomenological to mechanistic and predictive fundamental understanding of friction for crystalline materials, including engineering alloys.

## 1. Introduction

Contrary to the centuries-old conventional wisdom that unlubricated self-mated pure metal contacts cold-weld, gall and invariably produce high friction and wear [1–7], we present evidence that it is in fact possible to achieve low friction (with friction coefficients of $\mu < 0.5$) with pure bare metals such as pure Cu and Au. There are rare instances in the literature where this behavior has been shown before without surface modifying films, most prominently in Prasad et al. [8] for pure Ni and by Tamai for pure Cu and Au [9]. More interestingly, pure metals exhibit an apparent bi-stability in their friction behavior, where in contact conditions associated with low friction the trends in friction appear to be converging, while the trends diverge for higher friction. Various authors [8, 10–15] have reported that this unusual and highly desirable macro-scale behavior for pure metals and alloys are associated with the formation and persistence of a thin layer of highly refined, ultra-nanocrystalline (10 to 20 nm grain size) metal at the sliding surface, though the conditions and mechanism through which this layer is formed and how these two characteristics are correlated remains unclear. We propose a hypothesis and supporting evidence for a correlation between surface grain structure evolution and friction behavior of metals that is both predictive and quantitative, relying exclusively on materials properties and contact mechanics models to predict the bounds of observed friction regimes. The model is shown to accurately predict, without empiricism, the grain sizes of metal contacts in the low friction regime.

Central to the idea of a microstructural evolution model that drives metal friction behavior is the premise that the surface microstructure of a metal contact experiences a competition between grain refinement through plastic deformation and grain coarsening through grain boundary (GB) motion and dynamic recrystallization[16]. Be it through rolling[17, 18] or sliding[10, 11, 13, 14, 19, 20] there are many examples of grain refinement through cyclic contact via strain hardening[21], and more recently there are clear examples of coarsening of initially ultra-nanocrystalline metals[22, 23] as well as coarsening of more conventional fine-grained metal alloys[24, 25] toward an apparent stress-dependent equilibrium grain size. The idea that microstructural evolution and stability are the dominant characteristics of low friction behavior in bare metal sliding contacts is perhaps most clearly demonstrated by the electrical contacts community and literature in the form of noble metal alloys, primarily hard Au, as wear-resistant and low friction materials. With these alloys, the addition of as little as 0.1% by volume of an alloying species such as Ni or Co drastically reduces grain size[26] and imparts astonishing wear resistance and low friction at contact stresses that would otherwise grossly deform pure Au[27, 28]. While recent work has shown that it is possible to achieve low friction and wear as well as electrical contact resistances indicative of bare metallic junctions using non-noble metals such as pure Cu in non-oxidizing environments[12, 29–31], the advantages of relatively inert self-mated gold contacts as a baseline material are the reason why, in addition to Cu, it is one of the two material systems explored in this study.

As we show here, at sufficiently low applied stresses it is possible to achieve remarkably low friction ($\mu < 0.5$) with pure, coarse grained self-mated Au and even initially single crystal pure Cu, where severe galling and high friction ($\mu > 1$) would conventionally be expected to occur. While calling friction coefficients of $\mu < 0.5$ low may seem to be excessively generous considering superlubricious materials such as graphene ($\mu \sim 0.001$) [32–34], for pure, bare metals sliding in inert environments it defies convention that the friction behavior should be anywhere below values commensurate with strong and rapid interfacial bonding. For pure, bare metals, the characteristic low friction regime appears to be associated with the formation and persistence of a stable nanocrystalline (NC) surface film [8, 11, 19], with grain sizes well below 100 nm. Similar to the formation of a Beilby layer [35] during metal polishing, we find that there is a well-defined stress and temperature regime where this highly refined surface layer can be made to persist. We calculate the limits of this regime for Cu and Au self-mated contacts and experimentally validate the predicted limits, by systematically varying contact stresses and temperatures. We propose a generalized model for FCC metals that allows prediction of three distinct friction regime boundaries (persistent low, transient and immediately high friction) in the space of stress, temperature and time, and demonstrate its validity for two FCC metals with significantly different properties. This model does not require empirical fitting and relies exclusively on well-defined materials properties (e.g. stacking fault energy, Burgers vector, and shear modulus), fundamental plasticity models (i.e. Yamakov et al. and van Swygenhoven et al.)[36–39], and an appropriate contact mechanics model (e.g. Hamilton)[40, 41] used to determine the surface stress that drives grain size evolution. We chose the Hamilton contact model as there is a tractable and simple expression for the rapid calculation of maximum von Mises stress at the surface of a sphere-on-flat contact, which is shown to be a sufficiently accurate estimate of the stress driving surface grain growth in metal contacts that have been run-in.

As Johnson[17, 21] and Hamilton[40] describe, at friction coefficients $\mu > 0.3$ the location of maximum stress exists at the surface, which agrees well with the idea that these highly distorted ultra-nanocrystalline layers in metals should be present on the sliding surface for metal contacts. It is important to note that any calculation of maximum surface stress, such as that provided by Hamilton, is strongly dependent on the friction coefficient or surface shear stress. As a result, the determination of friction regimes as a function of surface stress, which is itself a function of friction coefficient, implies a feedback loop that can be either positive or negative. This feedback effect implies that the initial microstructure does not necessarily dictate the steady-state friction behavior of a material. This idea agrees well with extensive data showing that it is possible to achieve low friction with initially coarse bare metal at sufficiently low contact stresses, and that a range of contact stresses ought to exist where transient and periodic behavior is observed, one of the three regimes identified as part of this model and demonstrated through experiments here with Au and Cu. We illustrate this feedback loop of stress, friction coefficient and grain size in Fig. 1A and the corresponding microstructural evolution model in Fig. 1B.

## 2. Materials, Experimental Methods and Characterization

*2.1 Materials*

The Au-Au contact consisted of a 1.59 mm radius spherically-tipped bulk hard gold alloy pin (Deringer-Ney Inc., Neyoro-G alloy) brought into sliding contact with a high purity (> 99.999 %) bulk gold ingot. Details of the pin composition and preparation are described elsewhere[42]. The pure Au ingot was not polished to avoid transfer and embedding of hard particle media, instead it was pressed between polished W-C plates (average roughness, $R_a$ < 20 nm) in a hydraulic press to 3 ksi pressure to achieve a smooth surface finish, yielding an average roughness $R_a$ ~ 375 nm measured using a scanning white light interferometer (SWLI; Bruker Contour GT). The Cu-Cu contact consisted of a 1.19 mm radius UNS C10100 Cu ball bearing (McMaster-Carr) and a high purity (> 99.999 %) bulk single crystal substrate (Princeton Scientific, Princeton, NJ) with average surface roughness < 10 nm and an initial surface orientation of <100>.

*2.2 Friction Experimental Methods*

The instrument used for friction measurements is shown in Fig. 2 and functionally the same as the instrument described in a previous publication [13]. A brief summary of the instrument is given here. Orthogonal capacitance probes measured the deflection of a double-leaf spring cantilever (Anton Paar Standard and High Load cantilevers were used). Data was acquired at 100 Hz throughout sliding. Sample preparation is described in detail elsewhere[42], generally consisting of turning spherical tips on the Au alloy pins using a lathe, and polishing and rinsing with acetone then isopropyl alcohol prior to each experiment. The same rinse was used for all surfaces, though the Cu balls were dipped in an oxide reducing agent (bright dip) prior to testing. Fresh Au pins and Cu balls were used for each experiment. Flat substrates were fixtured with a small prescribed tilt designed to achieve a linear contact force ramp from 0 to 100 mN over a 10 mm long wear track. Sliding was unidirectional, with contact force linearly increasing during a cycle, at a constant 1 mm/s sliding speed. A pin or sphere was brought out of contact before returning to the starting position for the next cycle, also at 1 mm/s. For the data shown in Fig. 3A total of 1000 sliding cycles were used, except where noted. All friction experiments were performed in a glove box purged with dry nitrogen having < 10 ppm $O_2$ and < 40 ppm $H_2O$, measured using an Alpha Omega Instruments 3000 series oxygen analyzer and a Michell Instruments S8000 chilled mirror.

*2.3 Characterization of Grain Structure*

Characterization of the near-surface microstructure was done with a dual-beam FEI Helios Nanolab scanning electron microscope-focused ion beam (SEM-FIB). A trench was milled in the middle of wear tracks where high or low friction were measured, with the cross-sectioned plane oriented parallel to the sliding direction, and imaged using a secondary electron (SE), through-the-lens detector at 5 kV to produce the images shown in Fig. 4B and 4D. Top-down imaging was performed with an Everhart-Thornley SE detector also at 5 kV to produce images shown in Fig. 4A and 4C. Scanning transmission electron microscopy (STEM) was performed on an FEI Company Titan G2 80-200 operated at 200kV and equipped with a special aberration corrector on the probe-forming optics. Both bright-field and high-angle annular dark field images, shown in Fig. 4B and 4D, were acquired with a sub-0.136 nm probe with a current of 200pA and convergence angle of 18 mrad.

## 3. Experimental Evidence of a Low Friction Regime for Pure Metals

*3.1 Low Friction Regime for Pure Au in Dry Nitrogen*

Figure 3A shows friction coefficient data for a 99.999% pure bulk gold coupon sliding against a hard gold alloy pin for different values of applied normal force. Three regimes of friction behavior are found and characterized as having the following bounds: (regime I) steady-state low friction ($\mu < 0.4$) below a threshold normal force somewhere between 10 and 25 mN; (regime II) a time-dependent transition from low-to-high friction with periodic events likely due to wear that appear to restart the transient process; (regime III) steady-state high friction ($\mu > 0.8$) from the onset of sliding, above a threshold force between 75 and 100 mN. In Fig. 3B we show that this data cannot be collapsed by considering the evolution in friction behavior as a function of, approximately, accumulated plastic strain (i.e. average friction force multiplied by sliding cycles).

*3.2 Surface Grain Structure Analysis of Low and High Friction Cases for Au Sliding Contacts*

Fig. 4 shows representative surface SEM micrographs and FIB prepared wear track cross-section for both high and low ending-friction cases. These images show evidence of a nanocrystalline surface layer from an initially coarse grained pure Au substrate only in the case where low friction was observed, similar to those found in other literature reports of low friction with metals in various conditions [8, 10, 11, 43]. The low friction wear track was characterized (Figs. 4A-B) after an experiment where 10k cycles of unidirectional sliding were performed at a relatively low normal force of 1 mN, where friction behavior was consistently low throughout, $\mu \sim 0.3$. The high ending friction case (Figs 4C-D) consisted of 1k cycles of sliding at a higher contact force of 100 mN, where friction coefficients were consistently high, $\mu > 0.7$. Here, the surface grain size is significantly coarser and the nanocrystalline surface film is notably absent. As in all cases throughout this manuscript, all experiments were performed in dry nitrogen (< 10 ppm $O_2$ and < 10 ppm $H_2O$).

Transmission Kikuchi diffraction (TKD) maps are shown in Figs. 5A & 5C and high resolution TEM images in Figs. 5B & 5D, both on the same FIB-prepared cross-sectional liftouts from the 1 mN and 50 mN wear tracks for which data is presented in Fig. 2. These two cases correspond to well-defined wear tracks for self-mated unlubricated pure Au where friction (A) remained low at relatively low contact stress and (B) transitioned to high friction at higher contact stress. TKD showed relatively fine grained surface material in both cases, but only in the low friction case it appears that the near-surface microstructure was poorly indexed, indicating the presence of highly refined material present; with a 4 nm resolution, this TKD system only indexes grains that are about 4 pixels in size or larger (approx. grain size greater than 16 nm). In the lower normal force case we also see that the region of plastic deformation and grain refinement is significantly more surface-localized, as would be expected for a lower contact stress. The thickness of a NC surface film that is accommodating plastic strain via GB sliding would thus also be quite thin and on the order of a few 10s of nanometers, making it difficult to index via SEM-TKD (Fig. 5A). Further investigation of the same liftouts using high resolution TEM (Fig. 5B & 5D) was used to determine whether there was evidence of a nanocrystalline surface film. Bright field TEM images revealed that there was indeed an extremely fine grained layer (average grain size below about 20 nm) on the surface of the low friction track, and that the high friction surface was in fact larger in average grain size (with grains on the order of 20-50 nm). This explains our inability to resolve the surface grains in the low friction case using TKD and establishes that there is a critical albeit small difference in the average grain size for low and high friction cases.

## 4. MD Simulations Reveal Correlation between Grain Boundary Sliding and Friction

The experiments described above show that the friction between sliding metal contacts depends on the magnitude of the contact stress and how it affects the surface microstructure. Furthermore, experiments reported in the literature frequently point to another general trend at relatively high contact stresses; pure metals experience severe coarsening near the slider/substrate interface, while alloying, in contrast, can suppress grain growth and instead lead to refinement of the surface microstructure. The consequence is that friction can be lowered by alloying (e.g. hard gold). This observation was the motivation to understand the atomistic mechanisms responsible for the frictional behavior of metal contacts. In particular, we have performed molecular dynamics (MD) simulations to reveal the mechanisms leading to high friction in contacts between pure metals, as well as its reduction in alloys. We show that the different behavior in metals and alloys arises from the different sliding mechanisms accessible to pure and alloy systems. Specifically, in pure metals at relatively high stresses, stress-induced grain growth leads to shear along crystallographic slip planes, and sliding is dominated by dislocation-mediated processes. This corresponds to sliding between commensurate crystalline planes and results in high friction. In alloys, grains are refined near the sliding contact and shear occurs along GBs, leading to disordered contacts and reduced friction. The detailed understanding of the correlation between accessible sliding mechanisms and friction behavior described in this section -- achieved by contrasting MD simulations results from pure and alloyed metals -- stimulated the experimental investigations described above and were the inspiration for connecting friction regimes to surface microstructure and its evolution.

The experimental aspect of this work has concentrated on noble metals and non-noble metals sliding in inert gas environments in order to reduce complexities associated with oxidation. Although we discuss the role of oxidation in the tribological response of metals in the context of our generalized model, for both the experiments and the simulations, we will focus here on the response of pure metal in order to develop a generalized framework for the discussion and interpretation of our results. We have chosen to model Ag contacts instead of Au because of the availability of a well-tested atomic potential for both pure Ag and Ag-Cu alloys. We expect that, while results from simulations of Ag are interesting in their own right, they are also applicable to Au. This is primarily due to the similarities in the mechanical properties of the two materials. Ag and Au are both noble metals with low stacking-fault energies, and have the same FCC crystal structure and very similar lattice constants, bulk and shear moduli, hardness, and Poisson's ratios. We have also performed simulations on pure Au contacts and the results are nearly identical to those presented here for pure Ag. While the material systems explored in the simulations and experiments are different, the purpose of the simulations was solely to identify the differing atomistic mechanisms responsible for low and high friction, and as such, the results should be broadly applicable and not specific to a given choice of metals.

Our atomistic simulations use the embedded atom method (EAM), with the alloy potential for the Ag-Cu mixtures from Wu and Trinkle[44]. Although the experimental phase diagram shows that Ag and Cu are nearly immiscible at room temperature, this EAM potential has been optimized for surface diffusion of Cu islands on Ag, and predicts mixing of the bulk materials. It is therefore suitable for our purposes here. We created a nanocrystalline Ag substrate by melting a bulk FCC crystal of Ag at 1800 K and rapidly quenching it to 300 K over 100 ps. The resulting structure is intended to model an initially ultra-nanocrystalline tribological surface, with grains of approximately 5 nm in diameter and a strong {111} surface texture. Part of a 2-dimensional slice through the center of this 3-dimensional substrate is shown in Fig. 6A, where atoms locally in an FCC environment are colored according to their grain membership, those locally in an HCP environment (i.e., twins and stacking faults) are colored red, and those that cannot be classified as either FCC or HCP are colored black and represent GBs. The preferential alignment of twins and stacking faults parallel to the free surfaces at the top and bottom of the substrates indicate that the surface texture is dominated by {111} planes.

Our MD simulations were performed with the Large-scale Atomic/Molecular Massively Parallel Simulator (LAMMPS) code using a velocity-Verlet integration scheme with a time step of 1 fs. Our simulation geometry consists of matched pairs of polycrystalline slabs that are brought into contact by moving a rigid thin layer at the top of the upper slab, which will be called the loading layer, downward at a fixed velocity of 0.2 m/s. Shear is imposed by moving the loading layer laterally at a velocity of 2 m/s. Another thin layer of atoms at the bottom of the lower slab is also held rigid to allow for the control of compression and sliding. A layer of atoms adjacent to the bottom rigid layer in the substrate is used to control the temperature of the system at 300 K via a Langevin thermostat with damping time of 10 ps applied only in the direction orthogonal to compression and shear. We have calculated the temperature at the sliding interface and found only a minimal rise (~ 10 K). Each slab is initially 67 nm long in the shear direction, 34 nm high, and 17 nm deep (i.e., in the direction orthogonal to compression and shear).

To study the effects of alloying we randomly replaced 12.5% of Ag atoms in the previous system with Cu atoms to arrive at the composition of sterling silver. While Cu is soluble to only ~ 1% in Ag at room temperature[45], both the specifics of the EAM potential as well as the timescales accessible to our simulations obviate the segregation of the alloy into Ag-rich and Cu-rich phases. Experimentally, the effect of alloying is to introduce pinning sites in the microstructure, which lower GB mobility through solute drag and Zener pinning, effectively leading to a more stable microstructure. In our simulations the initial microstructure (specifically grain size and distribution) is identical in both the pure metal and the alloy, but the solute atoms still serve the purpose of preventing the coarsening of the grain structure, essentially enforcing a refined microstructure during sliding.

A snapshot of this sliding geometry is shown in Fig. 6 for both pure Ag (top images) and for Ag-Cu alloy (bottom images). The sliding interface is indicated upon first contact and after sliding with white squares. After sliding the principal sliding interface was identified by examining the velocity profile along the vertical *z* direction for both cases (Fig. 6C). In all instances, simulations were performed at constant normal forces.

Figure 7A shows the calculated shear stress for the Ag-Ag slab-on-slab contact shown in Fig. 6. The nanocrystalline Ag slabs undergo significant coarsening under shear and the sliding plane moves away from the initial contact plane as the shear proceeds. The coarsening is evident in Fig. 6 as well as Fig. 8 where we show the fraction of atoms that are identified as FCC, HCP or GB, based on their local arrangement. For pure Ag the increase of FCC atoms with sliding, with a concomitant lowering of GB atoms, indicates that grains are growing and consuming GBs.

For Ag-Cu alloys, the suppression of grain coalescence with subsequent refinement results in the sliding interface remaining near the initial contact between the two slabs (Fig. 6B). A quantitative measurement of the refinement is shown in Fig. 8, where for Ag-Cu as sliding progresses the number of FCC atoms decreases while the number of GB atoms increases. We take the mean calculated friction and compare pure Ag with Ag-Cu systems. Unlike with macro-scale experimental measurements of friction coefficient, where reported values are the ratio of measured forces, adhesion is a dominant factor in MD simulations sliding experiments[46, 47]. To remove the confounding effect of adhesion, friction coefficient measurements from atomistic simulations are therefore based on the slope of plots of normal and friction force. Simulation results are shown in Fig. 7A, with friction coefficient $\mu = 0.026$ for the alloy and $\mu = 0.075$ for the pure metal.

We can now discuss our hypothesis on the atomic-scale mechanisms behind the lower friction coefficient in alloys. In the Ag-Cu alloy, rather than a commensurate contact forming, grains remain predominantly distinct across the cold-welded slabs, and sliding is found to proceed along the new GBs formed at the junction. We also find evidence for significant grain refinement, both at the contact as well as in the bulk of the substrate. This is in contrast to pure Ag slabs, which coarsen to the point where effectively a bulk polycrystal is sheared against the loading layer rather than distinct slabs sliding against each other. We propose that dislocation-mediated processes are responsible for the high friction coefficient observed in pure Ag. In alloys the suppression of grain reorientation and growth, suppresses this sliding mechanism, and instead leads to GB mediated sliding and lower friction coefficient.

To demonstrate the validity of this hypothesis we further studied systems that not only allow the same contact geometry for both materials but also force the sliding mechanism to be identical. We then would expect similar friction coefficients in both cases. Demonstration of high friction from dislocation mediated plasticity in the alloy is achieved by switching to a fully miscible alloy of Ag with 10% Au. The EAM potential of the Ag-Au alloy is taken from Zhou et al. [48], which is generated from the EAM potential of Ag and Au via a mixing rule. While this potential has not been fully verified, we find that the friction for this alloy is high, and nearly the same as for both pure Ag and pure Au, indicating that it is adequate for our purposes here. We note that grain coarsening is also clear for this alloy in Fig. 8, where for the Ag-Au alloy the number of FCC atoms grows during sliding at the expense of GB atoms, similar to pure Ag. The growth in this system is slower than in the pure case, but clear nonetheless.

To confirm that GB sliding leads to low friction, we use a rigid plate of nanocrystalline Ag (or Ag-Cu) sliding against a thick mobile Ag (or Ag-Cu) substrate. Since rigid atoms in the plate suppress the grain coalescence and growth, sliding occurs along GBs in both the pure metal and the alloy. This case is discussed in detail below. Another option for suppressing the grain reorientation is to use rigid tips of a spherical-cap shape, modeling single asperity experiments in which the tip material is much harder than the substrate. We have performed these simulations as well and found the calculated µ in both cases to be near 1.0. Such high friction coefficients are due to the fact that the rigid tip leads to substantial plowing of and damage to the substrate material. A further motivation for the use of a rigid plate geometry is to avoid this plowing deformation.

The grain analysis and velocity profiles (not shown) demonstrate that for both the pure Ag and the Ag-Cu alloy, cold welding of the rigid plate to the substrate occurs, and a ~1 nm thick transfer film forms at the contact. Moving the rigid plate laterally at a constant velocity of 2 m/s leads to sliding predominantly along the boundary between the transfer film and the substrate. This boundary itself consists of multiple linked GBs where grains in the transfer film end and those in the substrate begin. As hypothesized, the rigid plate suppressed the grain coalescence even for pure Ag and the sliding occurs along GBs in both systems. The friction of the two systems is compared in Fig. 7B, with µ = 0.031 for the pure Ag and *µ* = 0.035 for the Ag-Cu alloy. Note that the friction coefficient is even slightly larger for the alloy in the rigid plate geometry. This is in sharp contrast to Fig. 7A where, for the same contact geometry, different sliding mechanisms in the pure Ag and the Ag-Cu alloy make the friction coefficient of the former larger by a factor of 3. It is also evident that in all cases where GB sliding dominated (i.e. both pure and alloyed for the rigid-on-mobile and only the alloyed case with mobile-on-mobile slabs) the friction coefficients were nearly identical.

The main simulation results are summarized as follows. In pure metals, cold welding, microstructural reorientation, and grain growth at the contact under shear leads to dislocation-mediated plasticity and sliding via FCC slip planes. The interface becomes a commensurate contact, which directly results in high friction. When the metal is alloyed, particularly with a material that has a different lattice constant (as is the case for Ag-Cu systems), the microstructural reorientation and growth is suppressed, and sliding occurs along GBs. The packing of atoms in GBs is essentially disordered, and therefore the friction is lowered. A similar mechanism of low friction resulting from GB rotation (as compared to high friction with dislocation controlled plasticity) has also been recently proposed to explain the different friction behavior observed in nanocrystalline Ni films[8].

**5. The Link Between Friction Behavior and Microstructural Evolution**

*5.1 MD Provides a Connection between Dominant Shear Mechanism and Friction Behavior*

In metallic contacts under shear, the microstructure at the contact is affected by a competition between grain refinement due to plastic deformation and coarsening due to stress-assisted grain growth. The simulations show that these mechanisms are directly tied to the tribological response. When applied stress is low, refinement dominates and this is associated with reduced friction and the prevalence of GB sliding as the primary mechanism of shear accommodation. At higher applied stresses, coarsening dominates and crystallographic slip is observed even in (initially) nanocrystalline metal contacts, and higher friction is observed. The ratio of friction coefficients between the two cases, one dominated by GB sliding and the other by crystallographic or intra-grain slip (Fig. 7A), was found to be about 1:3. While a direct comparison of friction coefficients between MD for pure and alloyed Ag (alloying used to suppress dislocation activity) and experimental results for Au or Cu is tempting, it would be unreasonable to do so given the inherent omission in MD of parameters such as surface roughness and contact curvature that impact friction behavior, not to mention the extreme difference in grain sizes between MD and experiments; it is obvious that direct comparisons between nanoscale simulations and micro/macro experiments are unreasonable. However, if the experimentally observed regimes are fundamentally linked to a transition between GB sliding and dislocation mediated plasticity at the metal interface, then it is reasonable to assume that at the extremes of high and low friction (regimes I and III in Fig. 3A) we should observe a ratio between the two friction coefficient extremes similar to the MD values. Referring to Fig. 3 again we see that the friction coefficient for Au varied between about 0.4 and 1.3, giving a ratio of about 1:3, remarkably close to the MD determined ratio. In fact, this 1:3 ratio also appears in literature reports documenting the low/high friction of metals, for example Padilla et al. [11] and Prasad et al. [10] work with nanocrystalline Ni.

Kinetic stabilization of GBs via Zener pinning [49] and solute drag [50], achieved through alloying as with hard gold coatings [26], influences this competition by limiting or effectively mitigating grain growth. Grain refinement then dominates even at relatively high stresses, and this enhanced refinement leads to the formation of nanocrystalline layers as seen in many low friction metal contacts [8, 11, 12, 42]. For pure metals, grain refinement from sliding contact of an initially coarse grained microstructure can happen rapidly as shown by Johnson [21], and this surface grain size refinement implies significant localized hardening and increased flow stress. This also implies that a single contact cycle at sufficiently low stress may generate the fine grained microstructure that is required for GB sliding and here associated with relatively low friction. At moderate stresses progressive cycles of contact will gradually coarsen the microstructure until the onset of dislocation mediated plasticity and high friction. At this point there are two possible outcomes, demonstrated in Fig. 3A for Au contacts: the process can reset through wear and begin a new low-to-high friction transition (regime II), or at sufficiently high surface stress it can lead to unrecoverable perpetually high friction (regime III) due to friction feedback enhancing surface stress beyond a critical stress.

Having established through experiments, microscopy and MD simulations that the friction regimes of metals correspond to a transition between two dominant plasticity mechanisms, we can now begin to define a friction regimes map where friction behavior is fundamentally a function of average surface grain size and applied stress. To establish predictive bounds of behavior we adapt the plasticity regimes model proposed by Yamakov et al.[39] with the additional inclusion of time-dependent grain size evolution. Fig. 9A is a recasting of Fig. 1B in the style of Yamakov et al.'s [39] plasticity regimes model with variables defined below.

Based on the evidence presented earlier that an initially coarse grained metal can be driven to a nanocrystalline surface grain structure and associated low friction, and that an initially ultra-nanocrystalline metal alloy can be driven through cyclic stress to coarser surface grain structure and high friction, we postulate that there must exist a stress-dependent equilibrium grain size toward which the surface grain size is driven in a tribological contact. The concept of an "equilibrium" grain size has been proposed and experimentally investigated by Derby[51] for metals during dynamic recrystallization and by De Bresser et al.[25] for ionic solids in the context of geological interfaces. Of greater relevance to the present work, Pougis et al.[24] extended Derby's work by investigating the concept of equilibrium grain size in metals exposed to severe plastic deformation, including a review of literature data for various pure metals, alloys and ionic solids. In a recent publication [23] the authors have shown that high thermal stability Ni-W alloy films exhibited a low stress friction transition correlated to a change in the average surface grain size. In this instance, surface grain size evolved from initially ultra-nanocrystalline (5 nm) to approximately 140 nm at high stress and high friction was found, but at sufficiently low contact stresses grain size evolved to about 30 nm and low friction persisted. We show in the ensuing discussion that 30 nm is below a critical value (of about 36 nm for Ni-W) that may be generally determined for metals, both pure and alloyed, based exclusively on materials properties.

*5.2 Calculating the Critical Grain Size for the Transition between GBS and Dislocation Mediated Plasticity*

While a tribological contact is a complex and dynamic system impacted by stochastic processes such as wear, we propose that a limiting grain size exists where plasticity models predict that it is no longer possible to support the inclusion of a stacking fault; for GB sliding and low friction to prevail the average surface grain size must be small enough [52] to mitigate the generation of stacking faults [37, 39, 52]. We propose that surface grain size under stress is driven towards the dislocation splitting distance $r_e$, which is related to the applied stress ($\sigma_a$) and defined in Yamkov et al. [39],

$$r_e = \frac{r_0}{1 - \sigma_a / \sigma_\infty} \qquad \text{Eq. 1}$$

where $r_0$ is the equilibrium (zero stress) dislocation splitting distance, $\sigma_a$ is an applied stress and $\sigma_\infty$ is the theoretical shear strength of the metal, or the stress at which the splitting distance becomes infinite [53],

$\sigma_\infty = \frac{2\gamma_{sf}}{b}$. It follows from Schmid's law that the stacking fault energy value used to compute the the theoretical shear strength should correspond to that of the slip system with the lowest energy barrier for slip; for FCC metals this corresponds to the close-packed {111} plane. The lowest of the intrinsic and extrinsic stacking fault energy should be used based on the same reasoning, that the lowest energy system will prevail in generating intra-grain slip to occur. Following Yamakov et al.[39], hereafter we utilize the reduced stress parameter, $\tilde{\sigma} = \frac{\sigma_a}{\sigma_\infty}$. Froseth et al.[53] define $r_0$ for a purely edge dislocation as,

$$r_0 = \frac{(2+\nu)Gb^2}{8\pi(1-\nu)\gamma_{sf}} \qquad \text{Eq. 2}$$

which is exclusively a function of materials properties: Poisson's ratio $\nu$, intrinsic stacking fault energy $\gamma_{sf}$, shear modulus $G$ and the magnitude of the Burgers vector $b$, of the corresponding dislocation.

We manipulate the definition of equilibrium splitting distance to give the stress as a function of grain size, and choose $r = 2r_0$ nm, *which defines the equilibrium grain size at which the onset of dislocation activity occurs*[39]. This is the value of $r$ at $\frac{1}{2}\sigma_\infty$, where Yamakov et al.[39] proposed that there is a cross-over from GB sliding to dislocation mediated plasticity along the line $d = r$.

We can now say that at stresses above the theoretical shear strength, corresponding to reduced stresses $\tilde{\sigma} \geq 1.0$, deformation is always accommodated via dislocation mediated plasticity and friction is therefore high. This bound indicates the onset of the stress regime in which a nanocrystalline surface structure is immediately unstable. For stresses below the theoretical shear strength, we postulate that the surface grain size is driven toward the equilibrium splitting distance. For reduced stresses in the range $0.5 < \tilde{\sigma} < 1.0$ the surface grain size is expected to be driven toward a grain size that is sufficiently large to accommodate dislocations and high friction. However, if the initial microstructure is nanocrystalline, or as Johnson[21] predicts for soft coarse grained metals, rapidly refined through significant plastic deformation that takes place in a single cycle of sliding, then we expect initially low friction with a gradual transition to high friction as a result of stress-assisted grain growth. This is the mechanism we propose is taking place at intermediate stresses, labelled as regime II in Fig. 3.

In other words, this analysis implies that there exists an intrinsic stability threshold for a material at $\tilde{\sigma} \sim 0.5$, corresponding to a predicted transition from GB to dislocation mediated plasticity. Again, this threshold is purely defined by fundamental material properties. Above this threshold surface grain size will be driven toward a size exceeding the crossover from GB to dislocation mediated plasticity (e.g. $d = 2r_0 \sim 17.4$ nm for Au). Below the threshold, grains are driven towards a size in which GB sliding is maintained, leading to the often observed nanocrystalline surface layer in metals with low friction coefficients. Fig. 9B presents the adapted Yamakov et al. model with values calculated for Au. The calculated parameters for a number of other metal systems are given in Table 1.

To complete the stress-dependent portion of the friction regimes map an appropriate way of calculating the applied stress must now be defined. Based on experimental evidence[10, 11, 54] and contact mechanics models[21, 40] including the present work we have shown that the nanocrystalline films associated with low friction in metals are highly surface localized. For conventional sphere-on-flat geometries such as those used here and in the literature we thus utilize the Hamilton contact model [40, 41] as it provides the following simple expression for the calculation of maximum von Mises stress at the surface of a sliding contact,

$$\sigma_{a,\max} = \frac{3F_n}{2\pi a^2}\left[\frac{1-2\nu}{3} + \frac{(4+\nu)}{8}\pi\mu\right]$$

The max surface stress is a function of the measured friction coefficient $\mu$ and normal force $F_n$, Poisson ratio $\nu$, and the purely elastic Hertzian model[40, 41] contact radius $a$.

We can now draw the stress-dependent bounds of friction behavior as a function of reduced stress with three distinct regimes: (I) persistent low friction below reduced stress $\tilde{\sigma} \leq 0.5$, (II) transient low-to-high friction behavior for reduced stresses in the range $0.5 < \tilde{\sigma} < 1.0$, and (III) immediate and persistent high friction at reduced stresses $\tilde{\sigma} \geq 1.0$. Based on experimental data we define low friction to be $\mu < 0.4$ and high friction to be $\mu > 0.8$.

A generalized reduced time parameter can be established based on a classical stress and temperature grain growth model. Following Dao et al. [55] and similar to Kumar et al.[36], the shear deformation rate in FCC metals has an exponential dependence on the applied stress, $\sigma_a$, and activation volume, $V^*$. We use this dependence, combined with a traditional grain growth model, to form a more general expression that accounts for temperature and stress activated grain growth, driving pressure due to GB curvature and the addition of an applied stress,

$$v_{gb} = \frac{4\gamma_{gb}}{d} M_0 \exp\left[\frac{-Q}{kT}\right] \exp\left[\frac{(\sigma - \frac{1}{2}\sigma_\infty)V^*(d)}{kT}\right] \quad \text{Eq. 3}$$

Note that the stress-related exponential has an asymptote at $\sigma_\infty/2$, indicating that this expression is only valid in regime II. We use this expression for GB speed, $v_{gb}$, to describe the rate of growth in the quasi-stable regime. Here, $\gamma_{gb}$ is the average GB energy (J/m$^2$), and $M_0 \exp\left[\frac{-Q}{kT}\right]$ the GB mobility (m/[s-Pa]), with pre-exponential mobility factor $M_o$, activation energy $Q$, Boltzmann's constant $k$, at surface temperature $T$ (K). As shown below, the boundary between low and high friction in the quasi-stable regime may be calculated by using the above equation to determine the time at a given stress required to grow grains to an average size $d = 2r_0$ assuming that gross plastic deformation of asperities during the first few cycles[21] leads to an average surface grain size of $d \sim r_0$, the smallest mechanically stable grain size[39]. Eq. 3 presents an expression for the rate of grain growth $\left(v_{gb} = \frac{\partial d}{\partial t}\right)$, so we integrate to get an expression for the time required to grow grains from $d = r_0$ to $2r_0$ as a function of stress. While a more complete model would allow activation volume to vary with grain size [55–57], this literature also suggests that for pure nanocrystalline metals it is reasonable to assume an approximately constant value of $V^* \cong 10b^3$; we have compared the time calculations from both models and found only a minimal difference, so we will use the simpler model here to avoid unnecessary complication.

We can now draw a general friction regimes map for metals as a function of reduced stress and time as shown in Fig. 10 that includes the transient behavior (regime II in Fig. 3). We have chosen to name the two identified critical stresses that define regime boundaries as the Beilby [35] and Hall-Petch limits, corresponding to reduced stresses of $\sigma/\sigma_\infty = 0.5$ and $\sigma/\sigma_\infty = 1.0$, respectively.

*5.3 Experimental Validation of the Predictive Model using Linearly Ramped Normal Force along the Wear Track Length (i.e. Stress and Time Dependent Friction Maps)*

In order to test the predictions of our model in a systematic and less time consuming way, we performed sliding experiments using self-mated Au and Cu contacts in unidirectional sliding with contact force that varied linearly along the length of the wear track. Resulting friction data as a function of cycles (i.e. contact time) and maximum surface von Mises stress is then plotted using the reduced stress and time parameters,

$$\sigma/\sigma_\infty = \frac{\sigma_{a,\max}}{\sigma_\infty} \quad \text{and} \quad t/\theta = \left(\frac{2\lambda_a}{v_s}\right)\left(\frac{4\gamma_{gb}M_0}{r_0^2}\right) \quad \text{Eqs. 5}$$

For relatively rough and evolving surfaces we do not know the effective contact length, so we chose to use the average Hertz contact diameter ($\lambda_a \sim a \sim 10\ \mu m$) as an order of magnitude estimate. Fig. 11A shows the raw data from sliding contact experiments with Au and Cu, and Fig. 11B shows the same data processed into coordinates of reduced stress and time. The dotted lines overlaid on the plots correspond to our predicted regime boundaries for these materials using values calculated from parameters in Table I.

As can be seen in this data, the predictions are remarkably accurate, especially considering that the model uses only average values for material properties to determine the time and stress regimes for a macroscale tribological contact. We also recognize that for the self-mated copper sliding contact, even at traces levels of oxygen in our testing environment (< 10 ppm), the growth of a monolayer of oxide is nearly instantaneous[58]. While the role of contaminants in general – including oxides – on the formation and stability of these highly refined surface layers is not entirely understood, it is clear that mechanisms such as Zener pinning of grain boundaries may be a contributing factor. However, the predicted behavior was demonstrated for both self-mated copper and a self-mated noble metal system, indicating that perhaps the contribution of oxides is negligible.

The proposed model also provides an explanation for the apparent speed dependent friction behavior observed for initially nanocrystalline pure Ni by Padilla et al.[11] and Prasad et al.[8]. Varying sliding speed changes the contact time per cycle, with higher speeds implying slower surface evolution, and thus more cycles required to effect the transition from GB to dislocation mediated plasticity. In the speed range of 0.1 to 10 mm/s, contact heating is negligible[22], removing it as a source of the speed dependence. Notable in their friction data was an apparent stress threshold below which low friction was always observed, in both cases corresponding to a Hamilton surface stress of about 500 to 600 MPa, consistent with a calculated reduced stress of $\sigma/\sigma_\infty = 0.5$ (see Table I). High resolution microscopy[8] of the low friction Ni surface revealed a 50 nm thick nanocrystalline surface film with an average grain size between 5 and 10 nm. These values corroborate the prediction that low friction will be associated with a surface grain size for Ni in the range $r_0 < d < 2r_0$; for Ni $r_0$ was calculated to be 4.9 (Table 1), thus the predicted average grain size of surface metal is in the range $4.9 < d < 9.8$ nm, in remarkable agreement with the measured values of 5 to 10 nm.

It is possible to identify a third critical stress that accounts for those material systems with relatively high stacking fault energies, and thus such small $r_0$ that the flow stress (i.e. hardness) of the metal is smaller than the predicted Beilby limit (at $\sigma/\sigma_\infty = 0.5$). Because $\sigma_\infty$ depends linearly on $\gamma_{sf}$, in some metals with sufficiently high SFE, the Beilby and Hall-Petch limits exist at stresses that exceed the bulk strength of the material. In these materials, gross deformation and a sharp, non-time dependent transition between low and high friction behavior is expected to occur at applied stresses approaching the flow stress or hardness, rather than above the Beilby limit. However, even for these materials it should be theoretically possible based on the proposed framework to apply gentle enough contact pressure to achieve low friction. Because the contact stress field typically extends deeper into the material than the relatively shallow depth of the nanocrystalline surface films responsible for low friction, it is possible that the Beilby limit would exist at stresses substantially higher than the bulk hardness, such that gross deformation through plowing would occur first.

One example of such a material is Al, having a SFE of 166 mJ/m$^2$ (Table I). Correspondingly, the Beilby limit is expected to exist at 560 MPa. From experimental micro-hardness test data in the literature[59], for a conservative range of grain size between 100 to 500 nm, the hardness was approximately in the range 350 to 700 MPa. We correlate measured hardness to uni-axial strength using the relationship established by Tabor (pp. 176-177 of [17]), where he showed that for an axisymmetric elasto-plastic contact of arbitrary shape hardness is about 2.8 times the yield or flow stress in simple compression. Applying the 2.8 factor suggested by Tabor to determine a range of yield strengths we find that the approximate applied (Hamilton) stress where we should see gross deformation or plowing is at about 150 MPa, or somewhere in the reduced stress range $0.1 < \sigma/\sigma_c < 0.2$. In other words, a non-time dependent transition should be found at significantly lower stress even for relatively fine grained pure Al, well below the Beilby limit at $\sigma/\sigma_c = 0.5$. Preliminary experiments have confirmed the existence of this additional limit, though this will be presented in a future publication.

Finally, it is insightful to again reconsider literature data [8] for pure, fine grained Ni in the context of this additional limit. We speculate that the relatively high yield strength of Ni, another metal with a relatively high SFE similar to that of Al (see Table 1), is what enabled those investigators to achieve low friction at stresses up to the Beilby limit. While more work is needed to verify this hypothesis, we propose that a comparison of the dimensionless ratio of shear modulus divided by half the critical shear strength, $\frac{Gb}{\gamma_{sf}}$, provides a simple and effective way to generally determine whether this limit of bulk deformation exists at applied stresses above or below the Beilby and Hall-Petch limits for Ni and other metals. If we calculate the corresponding values of the dimensionless ratio $\frac{Gb}{\gamma_{sf}}$ for Au, Cu, Al, and Ni using the parameters in Table 1, we find that only Al has a relatively low dimensionless stress ratio value of 47, while the others possess similar values in the range 148 to 173, even though SFE varies among them by nearly a factor of 3. We suggest that when $\frac{Gb}{\gamma_{sf}} < 100$ it is likely that the bulk deformation limit, and thus bulk strength, will be the defining limit of maximum applied stress for which it is possible to observe persistent low friction that is mediated by GB sliding along highly refined surface films. For $\frac{Gb}{\gamma_{sf}} > 100$ it follows that the Beilby and Hall-Petch limits will be the corresponding limiting applied stresses for which it will be possible to achieve persistent or periodic low friction. If the bulk deformation limit is exceeded, the location of the interface along which shear occurs would dip into the bulk, away from the surface. This explains why it is possible to achieve persistent low friction with Ni contacts at stresses up to the Beilby limit despite a relatively high SFE as shown in published experimental data [8, 11].

**Summary and Future Work**

The Hall-Petch and Beilby stability limits, defining the boundaries of friction regimes I and III, are effectively material properties influenced primarily by SFE, a parameter that may be modified via alloying [59, 60]. It follows that experimental determination of the bounds of regimes I and III for alloys and metal matrix composites may provide an indirect means of estimating SFE. The time-dependent behavior of materials in transient regime II, particularly the time required to transition from GB to dislocation mediated plasticity, also depends on materials properties. Here the behavior depends on a GB energy and mobility, while the latter can be modified by alloying to introduce kinetic stabilization mechanisms such as Zener drag [60–62]. The proposed model provides a correlation between the time-dependent evolution in friction behavior and the rate of microstructural evolution, providing an indirect means of characterizing average mobility and GB energy, for engineered alloys.

We presented evidence supporting the postulate that the average surface grain size in a sliding contact at relatively low stresses evolves toward a value equal to the equilibrium splitting distance, $d \to r_e$. If $r_e < 2r_0$ then low friction will prevail as refinement drives surface grain size into a regime of GB sliding, associated with the formation of a nanocrystalline surface film. If $r_e > 2r_0$ then the surface grains will evolve toward a regime where dislocation mediated plasticity dominates, and only metastable GB sliding may be achieved via engineered nanocrystallinity and reduced GB mobility. If the reduced stress is between 0.5 and 1.0 then a transient (and periodic) low-to-high friction may be observed. This is a result of stress-driven grain coarsening that is reset by wear and subsequent rapid grain refinement of freshly exposed coarse grained material. For a sufficiently stable alloy, wear will occur slowly but at a sufficiently high rate to prevent the surface structure from evolving into the dislocation mediated plasticity regime. This is, for example, perceived as persistent low friction in hard gold coated electrical contacts. A third limit exists that is a function of bulk grain size (hardness) and differs from the Hall-Petch limit that is associated with the characteristics and evolution of surface microstructure. This bulk deformation limit generally occurs at high reduced stresses, but for high SFE materials such as Al can exist below the Beilby limit, such that it is not possible to achieve the low friction regime at stresses up to $\sigma\% = 0.5$. We also note the compelling possibility that the proposed model and experimental method may also be used as a means of approximately determining the SFE of pure and alloyed metals.

While the Hamilton model was shown to be relatively accurate for the calculation of applied stress, it is likely that neglected considerations may be required to make accurate predictions for contact geometries sufficiently different from sphere-on-flat. For example, roughness was neglected in the current model, but this may be a stress-enhancer of sufficient magnitude to become a dominant parameter in the determination of maximum von Mises surface stress[63]. On the other hand, for micro to macro scale contacts at all but extremes of stress and sliding speed (i.e. temperature), it is well known that the size of the contact area is roughness limited and effectively proportional to the hardness [63]. Work by Robbins et al. [64], for example, may be helpful in establishing how the proposed model extends to nanoscale metal contacts where adhesion and other factors omitted in this study become relevant. For example, there are tantalizing implications for the design of longer life micro-electromechanical Ohmic switches that would benefit from such efforts. It may also be useful to apply the transient response expression to bulk materials under stress, for example in the analysis of grain size evolution in tensile tests or indentation after prolonged exposure to known quasi-static biaxial loads.

In a concurrent publication we report on the impact of temperature on friction behavior of copper in inert gas environments[65]. Specifically, we show that it is possible to recover low friction behavior with self-mated pure copper at stresses that produce room-temperature friction coefficients greater than one simply by reducing the temperature to cryogenic values.

**Acknowledgments**

We thank Prof. Greg Sawyer (U. Florida) for providing insightful critique of the proposed model and its description, Stephen Foiles (SNL) for enlightening discussions on determination of grain boundary and stacking fault energies via simulations and comparison with experimental values, Michael Dugger (SNL) and Somuri Prasad (SNL) for numerous helpful discussions about historical research connecting tribological behavior with microstructure and surface composition, Tim Furnish (SNL) for helpful comments about the stacking fault energy of alloys, Paul Kotula (SNL) for acquisition of STEM images, and Brendan Nation (SNL) for assistance with design of experiments and the acquisition of friction and wear data. The authors also acknowledge helpful discussions with Jorge Argibay about time-dependent multi-variate analysis. This work was supported by the Laboratory Directed Research and Development program at Sandia National Laboratories, a multi-program laboratory managed and operated by Sandia Corporation, a wholly owned subsidiary of Lockheed Martin Corporation, for the U.S. Department of Energy's National Nuclear Security Administration under contract DE-AC04-94AL85000.

*Conflict of Interest: The authors declare that they have no conflict of interest.***References**

**Figures**

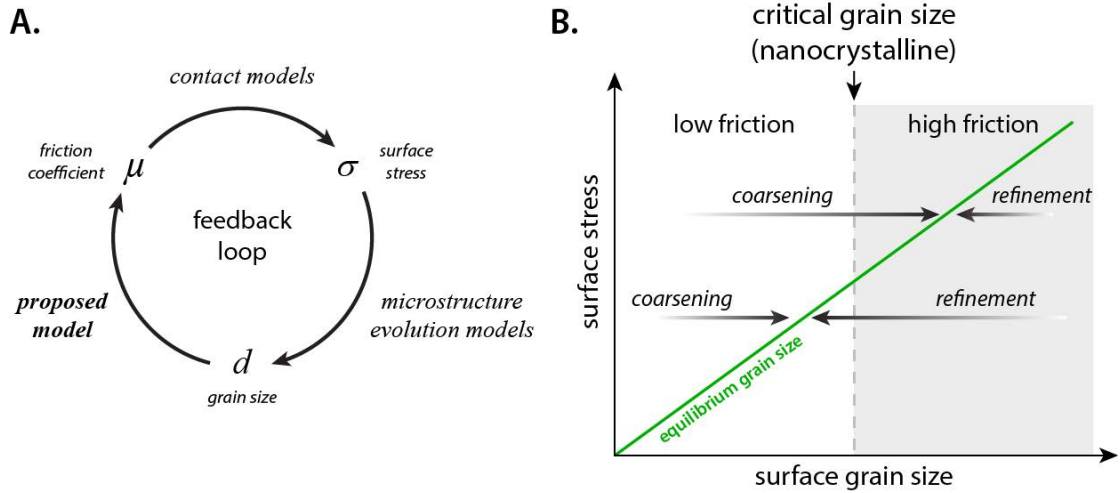

***Fig. 1.*** *(A) A diagram illustrating the feedback loop between maximum surface stress, σ, surface grain size, d, and friction coefficient, µ, and (B) the concept that there is a stress-dependent equilibrium grain size toward which surface grain size evolves, and a critical grain size differentiating between low and high friction behavior.*

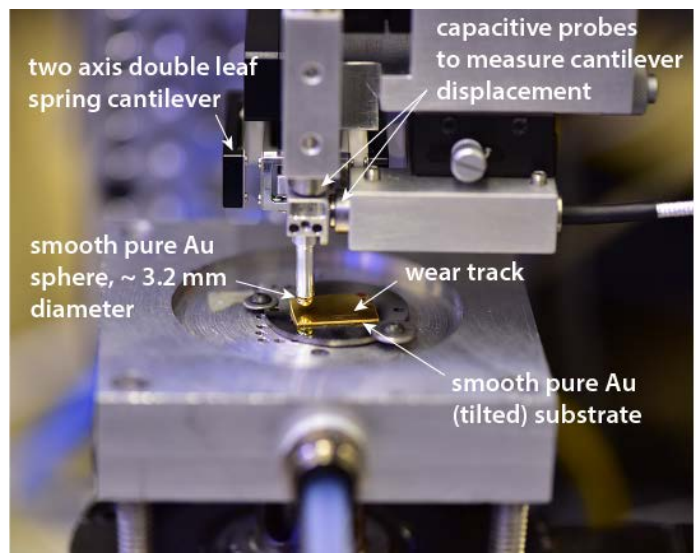

***Fig. 2.*** *Annotated photograph of tribometer showing a sliding experiment between a smooth pure Au ball and a pure Au substrate.*

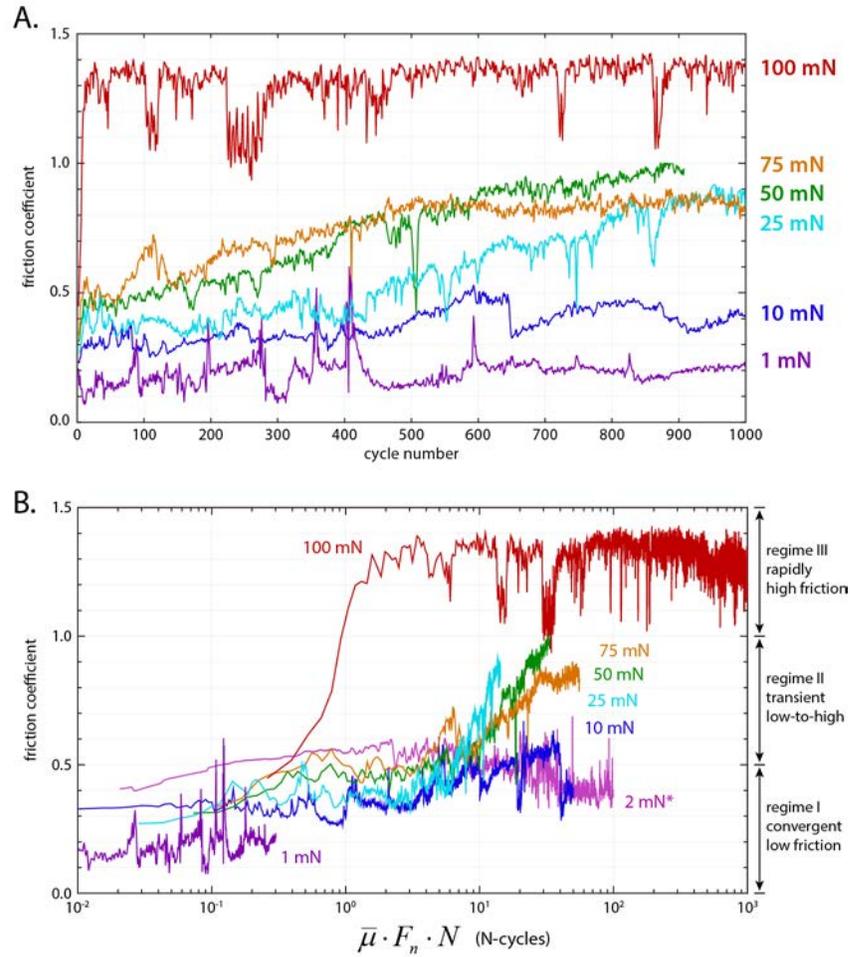

*Fig. 3.* Friction evolution data for a hard Au spherically tipped pin sliding against a flat pure Au coupon in dry nitrogen at constant normal forces as a function of (A) cycles and (B) the product of average friction coefficient for each experiment, normal force and cycles, analogous to accumulated plastic strain due to shear; the 2 mN data corresponds to a longer, 100k cycle experiment that was performed using uni-directional sliding in a rotating pin-on-disk tribometer.

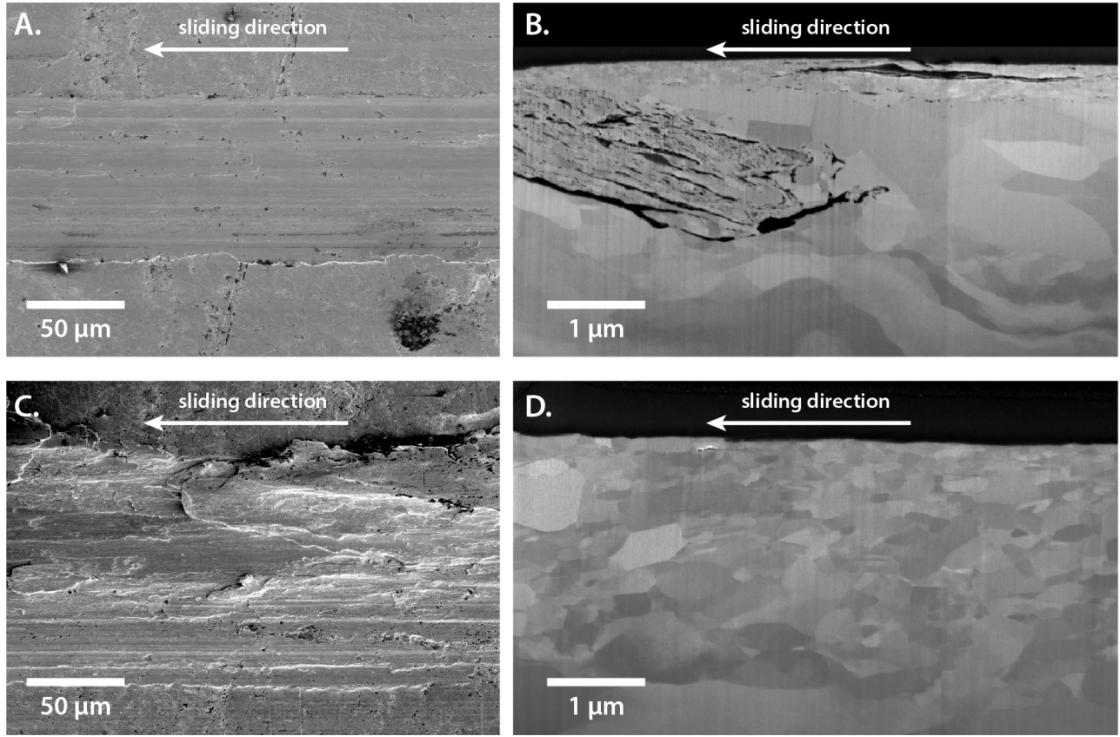

*Fig. 4.* Top-down SEM images of wear tracks on a bulk pure Au substrate representative of (A) low friction (μ ~ 0.3) and (C) high friction (μ > 0.7) steady-state sliding conditions against a hard Au spherically tipped pin, and representative FIB prepared cross-sectional SEM images from a location mid-track showing representative surface microstructures for the (B) low friction (nanocrystalline, d < 50 nm grains) and (D) high friction cases (coarse grained, d ~ 50 to 200 nm grains).

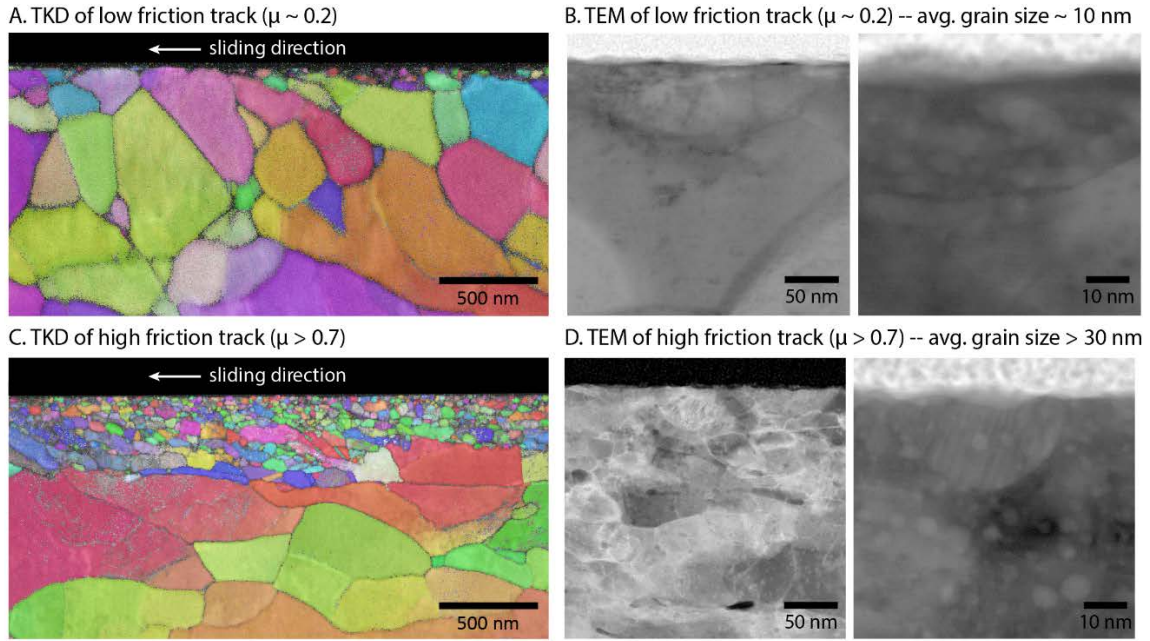

*Fig. 5.* (A and C) Grain orientation maps via TKD in SEM; color defines grain orientation. Low and high friction wear tracks correspond to experiments for which friction data is shown in Fig. 2, upon completion

*of (A) 10k cycles at 1 mN normal force and (B) 1k cycles at 50 mN. (B and D) Representative TEM micrographs of near surface grain structure for the same cross-sectional wear track liftouts shown in Fig. 4; a region of highly surface-localized nanocrystalline gold was found only in the low friction case.*

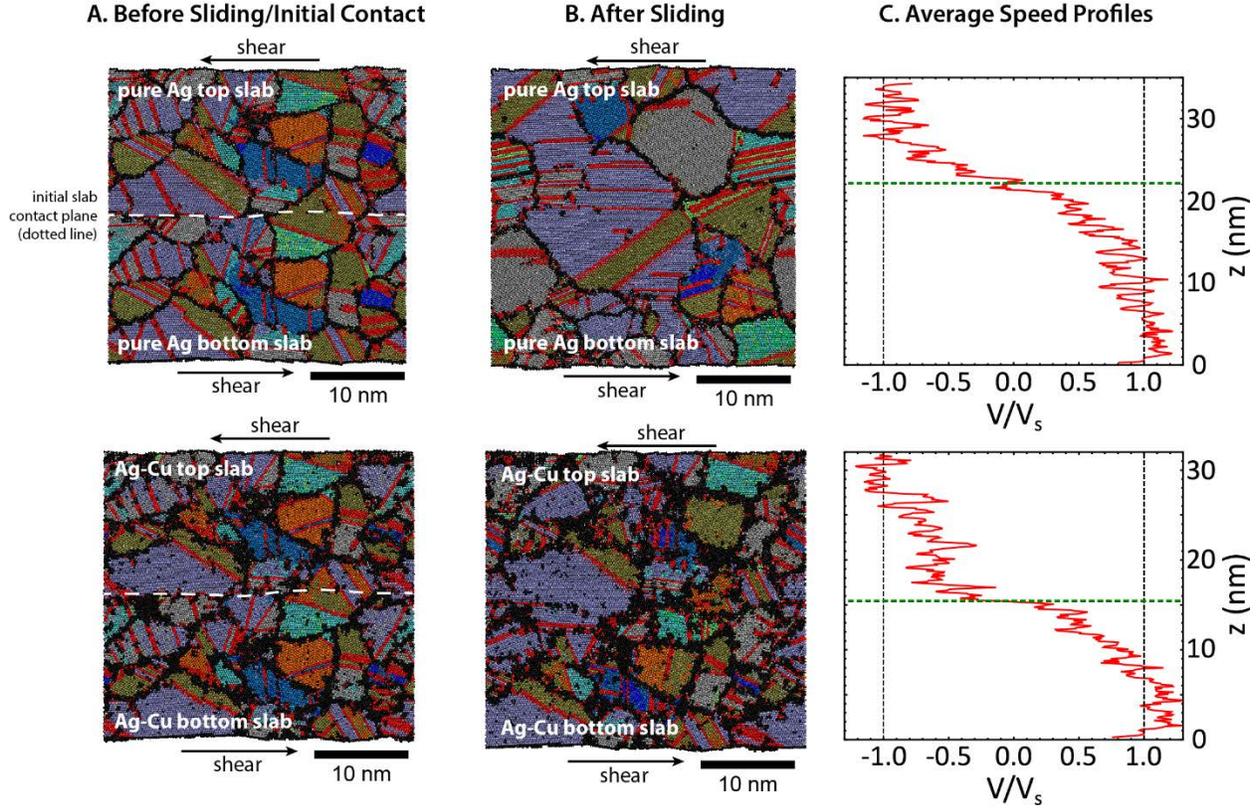

*Fig. 6. Snapshots of sliding contact simulations between two pure Ag and two Ag-Cu slabs at (A) initial contact before sliding and (B) after sliding. White dashed lines in (A) approximately indicate the location of the initial interface between the two slabs. In (C) we show the average velocity through the thickness in the sliding direction; the red lines are the velocities in the sliding direction.*

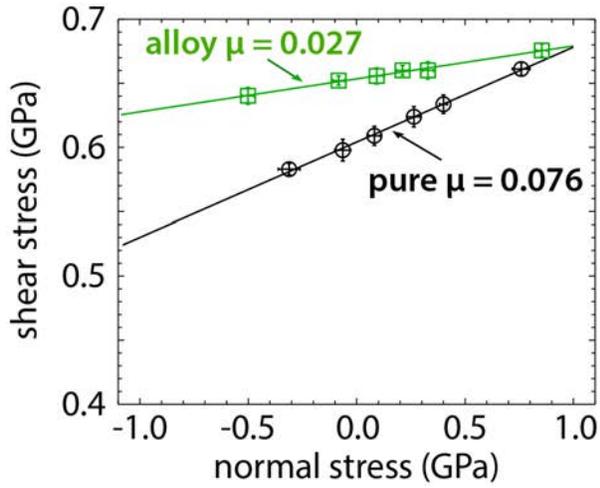
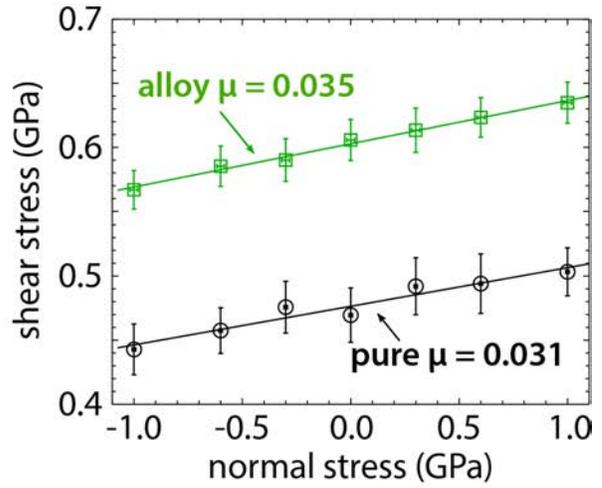

*Fig. 7.* *Friction vs. normal stresses for (A) the self-mated elastic slab-slab contacts of pure Ag (circles) and Ag-Cu alloy (squares) and (B) for rigid plate sliding over an elastic slab cases of pure Ag (circles) and Ag-Cu alloy (squares).*

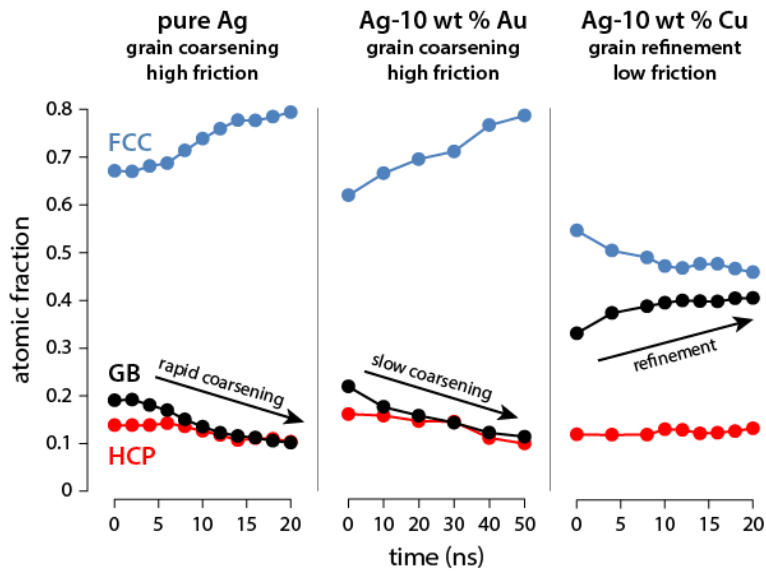

*Fig. 8.* *Evolution in grain boundary density for slabs of (left) pure, (middle) soluble alloy and (right) insoluble alloy after sliding at 300 MPa normal stress. Grain growth and high friction was found to occur in the pure and soluble alloy contacts where sliding occurs along FCC slip planes, but the immiscible alloy contact exhibited grain refinement (GB stability) and deformation was primarily accommodated via GB sliding with concomitant low friction.*

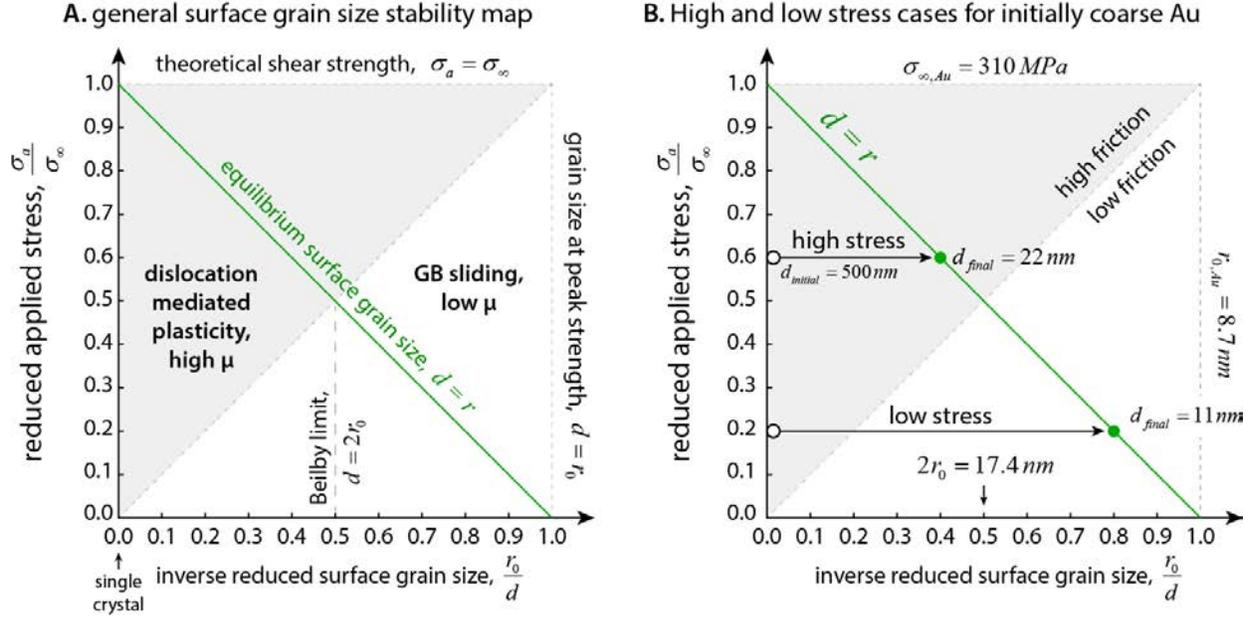

***Fig. 9. (A)*** *Friction regimes and dominant plasticity mechanisms map as a function of reduced stress and grain size, adapted and evolved from Yamakov et al.* [66]. *(B) Friction regimes and dominant plasticity mechanisms map showing two possible cases of time-dependent evolution of an Au-Au contact at low ( $\sigma_a/\sigma_\infty = 0.2$ ) and high ( $\sigma_a/\sigma_\infty = 0.6$ ) reduced stresses with an initial grain size of 500 nm.*

**Table 1.** Material properties and calculated parameters[67, 68].

| property | material system | | | | |
|---|---|---|---|---|---|
| | *Au* | *Cu* | *Al* | *Ni* | units |
| shear modulus, $G$ | 27 | 48 | 27 | 76 | GPa |
| Poisson ratio, $\nu$ | 0.44 | 0.36 | 0.35 | 0.31 | - |
| lattice constant, $a$ | 4.08 | 3.61 | 4.05 | 3.52 | Å |
| Burgers vector, $b$ | 2.88 | 2.55 | 2.86 | 2.49 | Å |
| SFE, $\gamma_{sf}$ | 45 | 78 | 166 | 128 | mJ/m$^2$ |
| GBE, $\gamma_{gb}$ | 378 | 625 | 324 | - | mJ/m$^2$ |
| HAGB mobility, $M_0$ | 3.84 x 10$^{-6}$ | 30 | 2 x 10$^{-2}$ | - | m/s-Pa |
| activation energy, $Q$ | 1.33 | 2.01 | 1.05 | - | x10$^{-19}$ J |
| **calculated parameters** | | | | | |
| *splitting distance*, $r_0$ | 8.7 | 5.9 | 2.0 | 4.9 | nm |
| critical grain size, $2r_0$ | 17.4 | 11.8 | 4.0 | 9.8 | nm |
| Hall-Petch limit, $\sigma_\infty$ | 312 | 611 | 1,117 | 1,029 | MPa |
| Beilby limit, $\sigma_a(2r_0)$ | 156 | 306 | 559 | 514 | MPa |

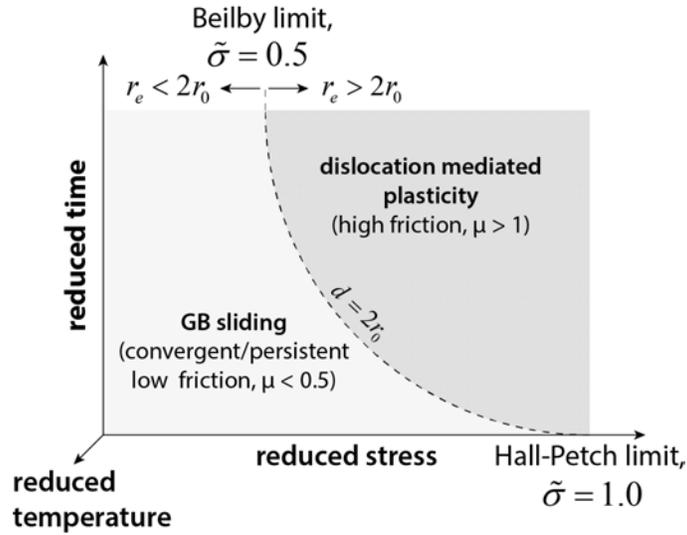

*Fig. 10.* *Generalized dimensionless friction regimes map for metals.*

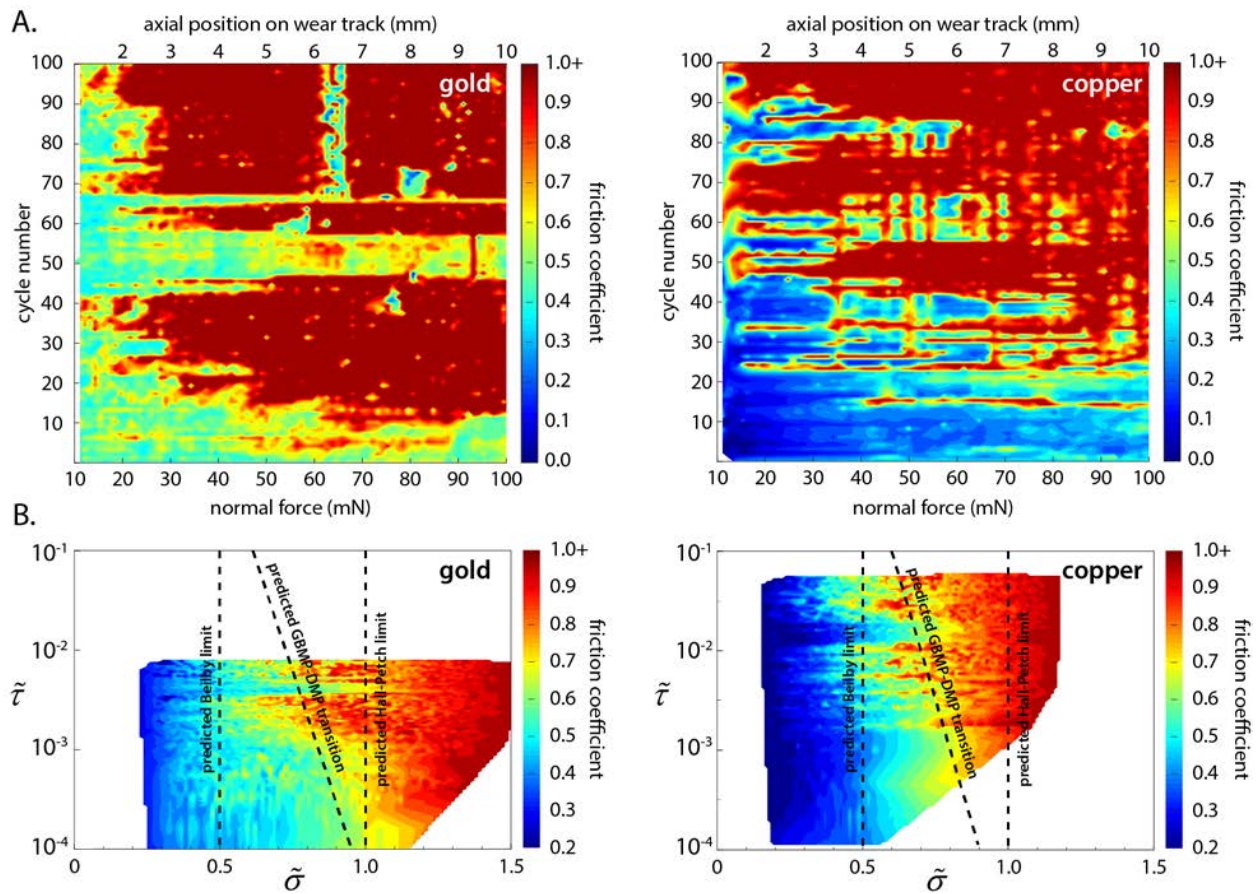

*Fig. 11.* *(A) Ramped normal force friction coefficient data for self-mated copper and gold contacts and (B) the same data processed into phase space maps of friction behavior for Au and Cu sliding contacts with calculated predictions of regime bounds overlaid using dotted lines.*

26